# Experimental demonstration of metamaterial "multiverse" in a ferrofluid

Igor I. Smolyaninov,[1,*] Bradley Yost,[2] Evan Bates,[2] and Vera N. Smolyaninova[2]

[1]*Department of Electrical and Computer Engineering, University of Maryland, College Park, MD 20742, USA*
[2]*Department of Physics Astronomy and Geosciences, Towson University,
8000 York Rd., Towson, MD 21252, USA*
[*]*smoly@umd.edu*

**Abstract:** Extraordinary light rays propagating inside a hyperbolic metamaterial look similar to particle world lines in a 2+1 dimensional Minkowski spacetime. Magnetic nanoparticles in a ferrofluid are known to form nanocolumns aligned along the magnetic field, so that a hyperbolic metamaterial may be formed at large enough nanoparticle concentration $n_H$. Here we investigate optical properties of such a metamaterial just below $n_H$. While on average such a metamaterial is elliptical, thermal fluctuations of nanoparticle concentration lead to transient formation of hyperbolic regions (3D Minkowski spacetimes) inside this metamaterial. Thus, thermal fluctuations in a ferrofluid look similar to creation and disappearance of individual Minkowski spacetimes (universes) in the cosmological multiverse. This theoretical picture is supported by experimental measurements of polarization-dependent optical transmission of a cobalt based ferrofluid at 1500 nm.

**OCIS codes:** (160.3918) Metamaterials; (160.4236) Nanomaterials.

## 1. Introduction

Recent advances in electromagnetic metamaterials and transformation optics gave rise to considerable progress in modeling unusual spacetime geometries, such as spacetime geometry near the big bang [1], black holes [2-4], wormholes [5], Alcubierre warp drive [6], spinning cosmic strings [7], Minkowski domain wall [8], and even metamaterial "multiverse" [9]. Hyperbolic metamaterials are especially interesting in this respect since extraordinary rays in a hyperbolic metamaterial behave as particle world lines in a three dimensional (2+1) Minkowski spacetime [1,10]. When this spacetime is "curved", metamaterial analogs of black holes [11] and the big bang [1] may be created. It appears that in a very strong magnetic field physical vacuum itself behaves as a hyperbolic metamaterial [12-14], so the metamaterial spacetime analogs appear to be quite meaningful.

While theoretical progress of this field is impressive, experimental developments appear to be considerably slower. Fabrication of 3D metamaterial structures required to accurately represent exotic spacetime models is extremely difficult, so experimental demonstrations are mostly limited to simplified 2D geometries [1,8,11]. Here we investigate a promising way to bypass these experimental difficulties by using ferrofluids. Magnetic nanoparticles in a ferrofluid are known to form nanocolumns aligned along the magnetic field [15], so that a wire array hyperbolic metamaterial may be formed at large enough magnetic nanoparticle concentration $n_H$. We investigate optical properties of such a metamaterial just below $n_H$. While on average such a metamaterial is elliptical, thermal fluctuations of the nanoparticle concentration lead to transient formation of hyperbolic regions inside this metamaterial. Extraordinary light rays inside these regions look similar to particle world lines in a 2+1 dimensional Minkowski spacetime [1]. Thus, thermal fluctuations in a ferrofluid give rise to transient "Minkowski spacetimes" which are somewhat analogous to individual Minkowski universes which appear and disappear as part of the larger cosmological multiverse [16]. This theoretical picture is supported by experimental measurements of polarization-dependent optical transmission of a cobalt based ferrofluid at 1500 nm.

## 2. Hyperbolic metamaterials as Minkowski spacetime analogs

As a first step, let us recall the analogy between extraordinary light propagation in hyperbolic metamaterials and world lines in 3D Minkowski spacetime, which is described in detail in refs. [1,10]. To better understand this analogy, let us start with a non-magnetic uniaxial anisotropic material with dielectric permittivities $\varepsilon_x=\varepsilon_y=\varepsilon_1$ and $\varepsilon_z=\varepsilon_2$. Any electromagnetic field propagating in this material can be expressed as a sum of ordinary and extraordinary contributions, each of these being a sum of an arbitrary number of plane waves polarized in the ordinary ($E_z \equiv 0$) and extraordinary ($E_z \neq 0$) directions. Let us define our "scalar" extraordinary wave function as $\varphi=E_z$ so that the ordinary portion of the electromagnetic field does not contribute to $\varphi$. Maxwell equations in the frequency domain results in the following wave equation for $\varphi_\omega$ if $\varepsilon_1$ and $\varepsilon_2$ are kept constant inside the metamaterial [1,10]:

$$\frac{\omega^2}{c^2}\varphi_\omega = -\frac{\partial^2\varphi_\omega}{\varepsilon_1 \partial z^2} - \frac{1}{\varepsilon_2}\left(\frac{\partial^2\varphi_\omega}{\partial x^2} + \frac{\partial^2\varphi_\omega}{\partial y^2}\right) \qquad (1)$$

While in ordinary elliptic anisotropic media both $\varepsilon_1$ and $\varepsilon_2$ are positive (corresponding to effective spacetime being Euclidean space), in hyperbolic metamaterials $\varepsilon_1$ and $\varepsilon_2$ have opposite signs. These metamaterials are typically composed of multilayer metal-dielectric or metal wire array structures, as shown in Fig. 1. Let us consider the case of constant $\varepsilon_1 > 0$ and $\varepsilon_2 < 0$, and assume that this behavior holds in some frequency range around $\omega = \omega_0$. Let us assume that the metamaterial is illuminated by coherent CW laser field at frequency $\omega_0$, and we study spatial distribution of the extraordinary field $\varphi_\omega$ at this frequency. Under these assumptions Eq. (1) coincides with the 3D Klein-Gordon equation describing a massive scalar field $\varphi_\omega$ in a 2+1 dimensional Minkowski spacetime. Note that the spatial coordinate $z = \tau$ behaves as a "timelike" variable in Eq. (1). When a metamaterial is built and illuminated with a coherent extraordinary CW laser beam at frequency $\omega = \omega_0$, the stationary pattern of light propagation inside the metamaterial represents a complete "history" of a toy (2+1) dimensional Minkowski spacetime. This "history" is written as a collection of particle world lines along the "timelike" $z$ coordinate. If adiabatic variations of $\varepsilon_1$ and $\varepsilon_2$ are allowed inside the metamaterial, world lines of massive particles in some well known curvilinear spacetimes can be emulated, including the world line behavior near the "beginning of time" at the moment of big bang [1]. Thus, mapping of monochromatic extraordinary light distribution in a hyperbolic metamaterial along some spatial direction may model the "flow of time" in an effective three dimensional (2+1) spacetime.

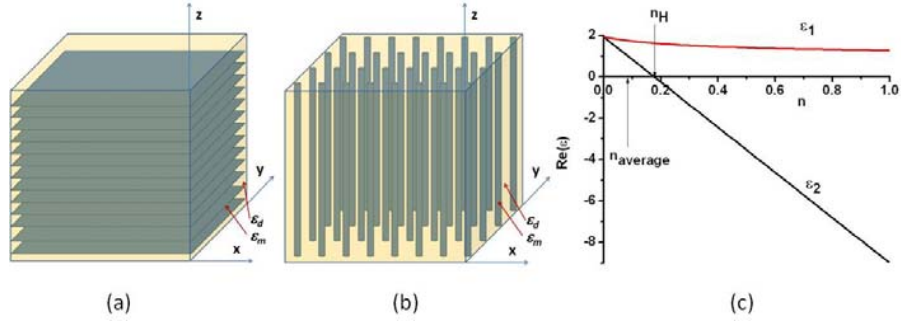

Fig. 1. Typical geometries of hyperbolic metamaterials: (a) multilayer metal-dielectric structure (b) metal wire array structure (c) effective medium parameters of the ferrofluid metamaterial.

## 3. Ferrofluids in external magnetic field as hyperbolic metamaterials

Ferrofluids are colloidal suspensions of nanoscale (<10 nm) ferromagnetic particles in a carrier fluid. Each ferromagnetic nanoparticle is coated with a surfactant to inhibit clumping. Due to this coating, magnetic attraction of nanoparticles becomes weak enough so that no particle agglomeration occurs in the absence of external magnetic field. On the other hand, when magnetic field is applied, ferromagnetic particles form nanocolumns, which are aligned along the magnetic field direction (see [15] and references therein). As a result, ferrofluid geometry becomes similar to metal wire array hyperbolic metamaterial structure shown in Fig. 1(b) with the column diameter approximately the same (10 nm) as nanoparticle size. Let us apply the standard metamaterial description to this medium. Diagonal components of the ferrofluid dielectric tensor may be obtained using Maxwell-Garnett approximation [17]:

$$\varepsilon_2 = \varepsilon_z = n\varepsilon_m + (1-n)\varepsilon_d \qquad (2)$$

$$\varepsilon_1 = \varepsilon_{x,y} = \frac{2n\varepsilon_m\varepsilon_d + (1-n)\varepsilon_d(\varepsilon_d + \varepsilon_m)}{(1-n)(\varepsilon_d + \varepsilon_m) + 2n\varepsilon_d} \qquad (3)$$

where $n$ is the average volume fraction of the ferromagnetic nanoparticle phase, and $\varepsilon_m$ and $\varepsilon_d$ are the dielectric permittivities of the ferromagnetic and liquid phase, respectively. We are interested in experimental situations which arise when $\varepsilon_m<0$, which correspond to ferromagnetic particles being metallic or being coated with metal. In such situations the ferrofluid becomes a hyperbolic metamaterial if

$$n > n_H = \frac{\varepsilon_d}{\varepsilon_d - \varepsilon_m} \qquad (4)$$

At this nanoparticle concentration $\varepsilon_2$ changes sign from positive to negative, while $\varepsilon_1$ remains positive if $-\varepsilon_m >> \varepsilon_d$. We are interested in optical properties of ferrofluids just below $n_H$. While on average such a ferrofluid remains usual "elliptical" material, thermal fluctuations of the nanoparticle concentration $n$ lead to transient formation of hyperbolic regions inside the ferrofluid. As described above, extraordinary light rays inside these regions look similar to particle world lines in a 2+1 dimensional Minkowski spacetime [1]. Thus, thermal fluctuations in a ferrofluid give rise to transient "Minkowski spacetimes" which are somewhat analogous to individual Minkowski universes which appear and disappear as part of the larger cosmological multiverse [16]. According to the concept of chaotic inflation, an infinite multiverse must contain Hubble volumes realizing all kinds of physical laws and initial conditions. In particular, an infinite multiverse will contain an infinite number of 3+1 dimensional Minkowski spacetimes (Hubble volumes), some of them being virtually identical to ours. It appears that fluctuating ferrofluids exhibit some similarity to this theoretical picture, albeit on a smaller scale and in smaller number of spacetime dimensions: inside the ferrofluid fluctuating 2+1D Minkowski spacetimes appear and disappear inside a larger 3D Euclidean space.

## 4. Thermal fluctuations of the nanoparticle volume fraction in ferrofluids

Let us evaluate thermal fluctuations of the dielectric tensor in ferrofluids. In principle, both $\varepsilon_m$ and $\varepsilon_d$ experience thermal fluctuations due to thermal fluctuations of metal and liquid densities. However, density fluctuations in liquids are proportional to $(\partial V/\partial P)_T$ [18], which is very small in a typical (incompressible) liquid far from its critical temperature. Thus, fluctuations of $\varepsilon_1$ and $\varepsilon_2$ must be dominated by thermal fluctuations of the nanoparticle volume fraction $n$ (see Eqs. (2) and (3)). If $N$ is the number of nanoparticles in a given volume $V$ of the ferrofluid, its standard deviation due to thermal fluctuations is [18]

$$\langle (\Delta N)^2 \rangle = N \qquad (5)$$

Thus, standard deviation of the nanoparticle volume fraction due to thermal fluctuations is

$$\langle (\Delta n)^2 \rangle^{1/2} = \frac{n^{1/2} v^{1/2}}{V^{1/2}} \qquad (6)$$

where $v$ is the volume of individual nanoparticle. In order for the macroscopically averaged "metamaterial" description to be valid, we need $V >> v$. Assuming $V > 10v$ limitation, the range of acceptable volume fraction fluctuations is $\langle (\Delta n)^2 \rangle^{1/2} \leq n^{1/2}/3$. At this fluctuation level the metamaterial description remains applicable. This consideration demonstrates that it is possible to choose a ferrofluid having average $n < n_H$, so that on average this ferrofluid will be a usual "elliptical" material, while considerable fraction of its volume will behave as a hyperbolic metamaterial due to thermal fluctuations of $n$. The local value of $n$ in the hyperbolic areas may temporarily exceed $n_H$ due to thermal fluctuations. We should also note that characteristic time scale of these fluctuations is much larger than the inverse light frequency at 1500 nm. Therefore, macroscopic electrodynamics description of these areas as hyperbolic metamaterials remains valid.

## 5. Experimental sample and setup

For our experiments we have chosen cobalt magnetic fluid 27-0001 from Strem Chemicals composed of 10 nm cobalt nanoparticles in kerosene with AOT (sodium dioctylsulfosuccinate) and LP4 (a fatty acid condensation polymer). The average volume fraction of cobalt nanoparticles in this ferrofluid is 8.2%, which is below $n_H$ at 1500 nm light wavelength: the value of $n_H=17\%$ can be calculated using Eq. (4) based on the dielectric constants of kerosene $\varepsilon_d=1.93$ and cobalt $Re\varepsilon_m=-9.0$ at 1500 nm [19]. Effective medium parameters of the ferrofluid in external magnetic field calculated using Eqs. (2,3) are plotted in Fig. 1(c). According to Eq. (6), for a (50 nm)$^3$ volume of the ferrofluid $\langle(\Delta n)^2\rangle^{1/2}=0.04$. Thus, at any given time 2.3% of such volume elements should have the local value of $n$ above $n_H$ (this corresponds to $\sim 2\sigma$ deviation from the average nanoparticle concentration). A similar estimate for a (30 nm)$^3$ volume of the ferrofluid produces $\langle(\Delta n)^2\rangle^{1/2}=0.08$ so that 16% of such volume elements should have the local value of $n$ above $n_H$ (this corresponds to $\sim 1\sigma$ deviation). These volume elements should exhibit transient hyperbolic behavior. The cutoff of the hyperbolic dispersion law has been analyzed in ref. [10]. It is defined by the metamaterial lattice constant. In our case, 17% volume concentration of nanoparticles results in average inter-particle distance of 18 nm. Thus, description of (50nm)$^3$ volumes as volumes occupied by a hyperbolic metamaterial does make sense. Typical instantaneous distribution of fluctuating hyperbolic regions in our sample calculated using Eqs. (6) and (2) is presented in Fig. 2. The instantaneous *xy* dependence of the particle number was calculated using a random number generator based on the standard deviation given by Eq. (5). If the local instantaneous value of $n(x,y)$ exceeded $n_H$, the dispersion law is locally hyperbolic. These hyperbolic regions are shown in red. As described above, hyperbolic regions in a ferrofluid shown in Fig. 2 behave as transient 2+1 dimensional Minkowski spacetimes, which temporarily appear and disappear inside a larger metamaterial "multiverse".

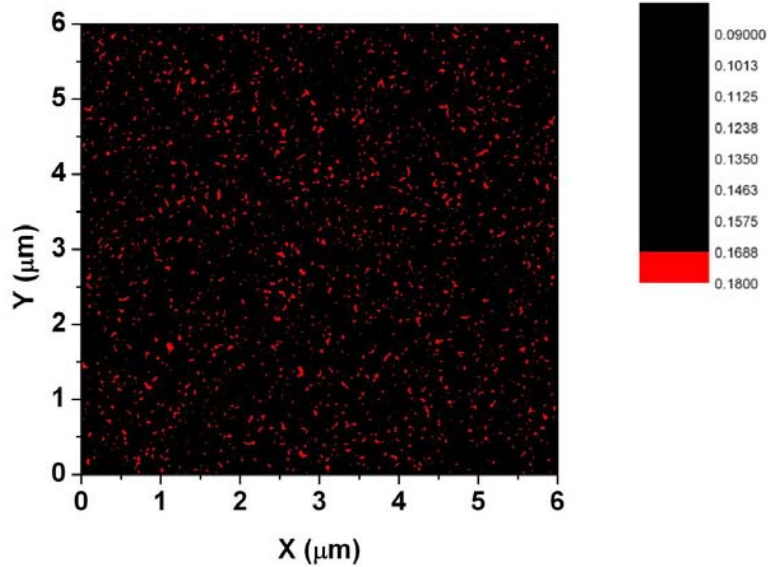

Fig. 2. Typical instantaneous distribution of fluctuating hyperbolic regions in the cobalt-based ferrofluid sample calculated using Eq. (6). Hyperbolic regions are shown in red. These regions behave as transient 2+1 dimensional Minkowski spacetimes which temporarily appear and diappear inside a larger metamaterial "multiverse".

Note that ferrofluids based on gold and silver-coated ferromagnetic nanoparticles proposed in [15] would require considerably smaller values of *n~2%*, and therefore would exhibit much smaller losses and much more pronounced hyperbolic behavior at 1500 nm. Therefore, similar experiments with gold and silver-coated ferromagnetic nanoparticles may be conducted much closer to the critical value $n_H$, leading to considerably larger size of transient hyperbolic regions inside the ferrofluid.

We have examined polarization-dependent optical transmission of the cobalt based ferrofluid at 1500 nm as a function of external magnetic field. Our experimental setup is shown schematically in Fig. 3(a). Linear polarized light from a 1500 nm laser has been sent onto the ferrofluid sample via a λ/4 plate, which made the illuminating light circular polarized. The cobalt based ferrofluid was placed in a 10 μm thick optical cuvette located between the poles of a magnet. Polarization state of the transmitted light has been analyzed using a polarizer as a function of external magnetic field. Temporal fluctuations of the transmitted signal have been also evaluated as a function of applied magnetic field.

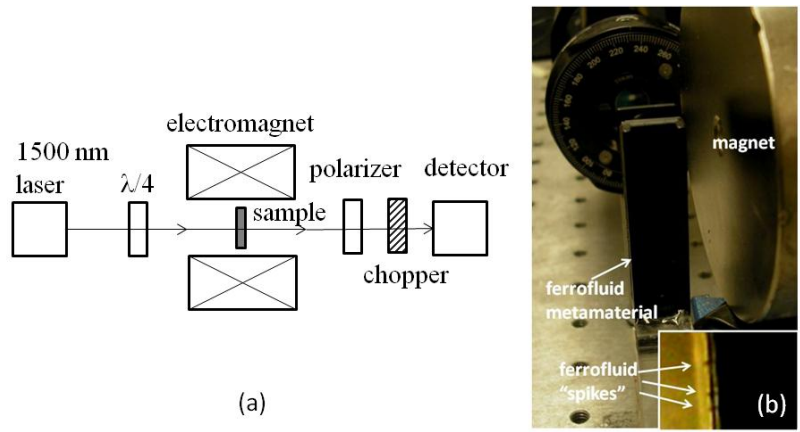

Fig. 3. (a) Schematic view of our experimental setup. (b) Photo of the ferrofluid metamaterial sample next to a permanent magnet. The inset shows excessive ferrofluid on the side of the cuvette, which forms "spikes" along the applied magnetic field.

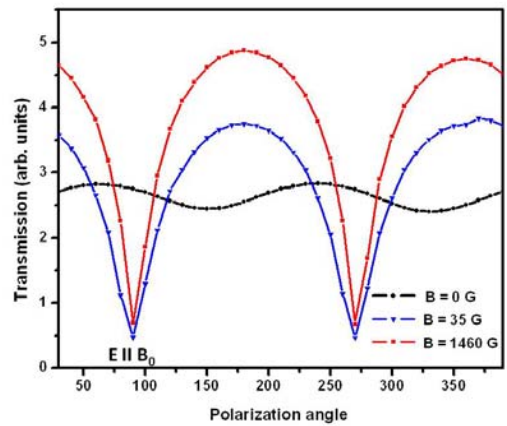

Fig. 4. Experimentally measured transmission of the cobalt based ferrofluid as a function of external magnetic field and polarization angle. Transmission signal was averaged over 2 minutes.

We should emphasize that unlike typical 3D hyperbolic metamaterial fabrication techniques described in the literature (see for example [20]), preparation of sample shown in Fig. 3(b) requires much less effort.

## 6. Experimental results

Experimentally measured transmission of the cobalt-based ferrofluid for three different values of the external magnetic field $B_0$ is plotted in Fig. 4 as a function of polarization angle ($\alpha=0^o$ corresponds roughly to $E$ field of the electromagnetic wave being perpendicular to $B_0$). While at zero magnetic field 1500 nm light transmission through the ferrofluid is isotropic (weak polarization dependence in zero field observed in Fig. 4 may be attributed to the quartz cuvette), it becomes strongly anisotropic when magnetic field is applied to the ferrofluid. 1500 nm light transmission exhibits pronounced minima when $E$ field of the wave is parallel to the direction of external magnetic field. This behavior is natural and expected if cobalt nanocolumns are indeed formed inside the ferrofluid. On the other hand, nonzero light transmission in these minima indicates that the ferrofluid at $n<n_H$ remains an elliptic material. According to our estimates, a 10 μm thick layer of hyperbolic metamaterial would have zero transmission when E field is parallel to the metallic nanocolumns.

Temporal fluctuations of the transmitted signal have been also evaluated as a function of applied magnetic field and light polarization as shown in Fig. 5. These measurements reveal considerable increase in temporal fluctuations of the transmitted signal in applied DC magnetic field. These results are consistent with the expected strong fluctuations of the dielectric tensor of the ferrofluid metamaterial, which is described in Section 4 above.

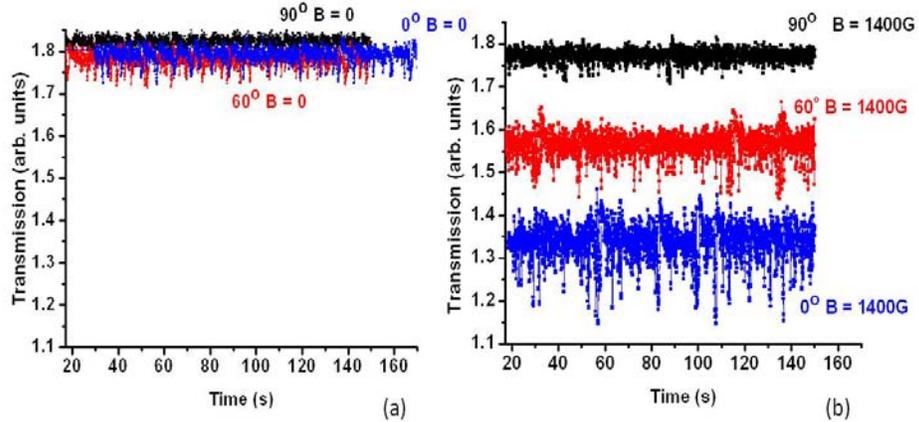

Fig. 5. Measured temporal fluctuations of the sample transmission as a function of light polarization and applied magnetic field: (a) measurements in zero field, (b) measurements in $B_0$=1400G. Angle between $B_0$ and $E$ is indicated for each time dependency.

It appears from Fig. 5 that in external magnetic field temporal fluctuations exhibit strong dependence on the polarization state of incident light. They are the strongest when $E$ field of the incident linear polarized light is directed along the external magnetic field. In addition to general increase in temporal fluctuations as a function of external magnetic field, short bursts of increased fluctuations are also observed. This behavior may be explained by sudden breakdown of nanocolumn ordering followed by nanocolumn rearrangement.

It is interesting to note that the observed strong polarization dependence of transmission fluctuations disappears at much lower concentrations of cobalt nanoparticles in the ferrofluid. Normalized fluctuations of optical transmission measured for different polarization states of 1500 nm light at 1% and 8.2% volume concentrations of cobalt nanoparticles are compared in

Fig. 6. This figure demonstrates that the observed effect indeed disappears far from the hyperbolic edge.

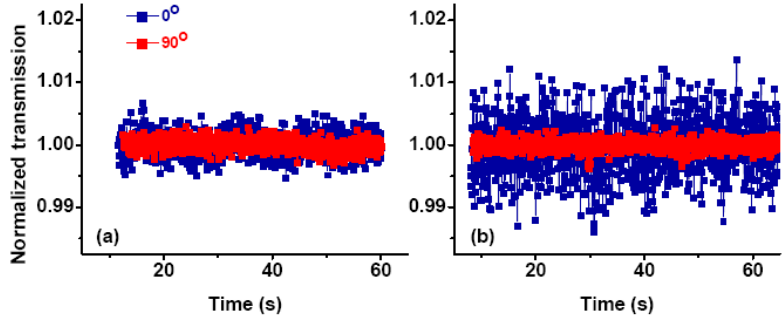

Fig. 6. Temporal fluctuations of normalized transmission in applied magnetic field at different concentrations of cobalt nanoparticles in the ferrofluid: (a) measured fluctuations as a function of light polarization at 1% volume concentration of cobalt nanoparticles, (b) measured fluctuations as a function of light polarization at 8.2% volume concentration of cobalt nanoparticles.

## 7. Conclusion

We have investigated optical properties of ferrofluids in the "sub-hyperbolic" $n<n_H$ range of volume fractions of metallic ferromagnetic nanoparticles. Polarization-dependent optical transmission of such a cobalt based ferrofluid at 1500 nm has been studied as a function of applied magnetic field. While on average such ferrofluid metamaterials are elliptical, thermal fluctuations of the nanoparticle concentration lead to transient formation of hyperbolic regions inside this metamaterial. These regions behave as transient 2+1 dimensional Minkowski spacetimes which temporarily appear and disappear inside a larger metamaterial "multiverse". The described experimental system may also be studied as yet another example of random hyperbolic medium [21]. As demonstrated recently in ref. [21], study of light propagation through such media may give important insights into electromagnetic properties of our universe immediately after the electro-weak phase transition leading to new estimates on the magnitude of magnetic fields which existed in the early universe.

**Acknowledgments**

This work is supported by the NSF grant DMR-1104676. We are grateful to J. Klupt for experimental help.